\documentclass[12pt]{article}
\usepackage{graphicx}
\usepackage[T2A]{fontenc}
\usepackage[cp866]{inputenc}
\begin{document}
\title{SUPERCONDUCTING PLATE IN TRANSVERSE MAGNETIC FIELD: NEW STATE}
\author{E.G. Batyev\thanks{e-mail:batyev@isp.nsc.ru} \\A.V. Rzhanov Institute of Semiconductor
Physics SB RAS, \\630090, Novosibirsk, Russia.}

\maketitle

\begin{abstract}A model to describe Cooper pairs near the transition point
(on temperature and magnetic field), when the distance between them
is big compared to their sizes, is proposed. A superconducting plate
whose thickness is less than the pair size in the transverse
magnetic field near the critical value $H_{c2}$ is considered as an
application of the model. A new state that is energetically more
favourable than that of Abrikosov vortex state within an interval
near the transition point was obtained. The system's wave function
in this state looks like that of Laughlin's having been used in
fractional quantum Hall effect (naturally, in our case -- for Cooper
pairs as Bose--particles) and it corresponds to homogeneous
incompressible liquid. The state energy is proportional to the first
power of value $(1 - H/H_{c2})$, unlike the vortex state energy
having this value squared. The interval of the new state existence
is greater for dirty specimens.\end{abstract} PACS {71.27.+a}

\section{Introduction}

  The difference of free energies of superconducting $F_S$ and normal $F_N$ states
 in Ginzburg--Landau theory is written as: \begin{eqnarray} \label{1} F_S - F_N \rightarrow
\int d{\bf r} \biggl\{\frac{1}{2M}
\Bigl|(\nabla - i\frac{2e}{c} {\bf A})\Psi({\bf r})\Bigr|^2 + \\
\nonumber \alpha \Bigl|\Psi \Bigr|^2 + \frac{\beta}{2}\Bigl|\Psi
\Bigr|^4 \biggr\}\ .\end{eqnarray} Only the part explicitly
dependent on the order parameter $\Psi({\bf r})$ is considered here.

Value $\Psi({\bf r})$ can be treated as the wave function of Cooper
pair (precise to the coefficient). All pairs are in the same state
(they form Bose--condensate) and, therefore, a function from one
coordinate is sufficient. Such an interpretation enables us to go
further and describe the system of Cooper pairs with the help of
Hamiltonian: \begin{eqnarray} \nonumber \textsf{H} =  \int d{\bf
r}\biggl\{\frac{-1}{2M} \Psi^+({\bf r}) \Bigl(\nabla - i\frac{2e}{c}
{\bf A}({\bf r})\Bigr)^2\Psi({\bf r}) +\\ \alpha \Psi^+({\bf
r})\Psi({\bf r}) + \label{2} \frac{\beta}{2} \Psi^+({\bf r})
\Psi^+({\bf r}) \Psi({\bf r})\Psi({\bf r})\biggr\}\ .\end{eqnarray}
Here $\Psi({\bf r}),\ \Psi^+({\bf r})$ are Bose--type operators in
the secondary quantization. If we replace them by simply functions
(to describe Bose--condensate), then there will be expression (1).

Such a generalization of GL theory was suggested many years ago in
contribution [1] but with no success. In the present study it is
shown that, in some cases, this approach leads to a new state which
is energetically more favourable than that following from a usual
formulation of GL theory. Namely, a new state turns up in a
quasi--2D case (superconducting plate) in the transverse magnetic
field near the upper critical field $H_{c2}$. Seemingly, it can be
proved experimentally.

First on the speculations in favor of a new approach with the use of
operator (2). 1) In the work [2], a phase transition in
superconductor was considered. The main thing shown in this study is
that, near the transition point, the diagram technique for a
singular part of two--particle Green function is the same as for
Bose--particles system. This singular part just describes the Cooper
pair. 2) It is possible to show that precisely the same ratios for
coefficients of GL theory, earlier obtained by Gor'kov (see e.g.
[3]), follow from the diagram approach used in [2] for the
two-particle Green function. That is, first, we compare
Bose--Hamiltonian -- for which there appears the same diagram
technique as for the two-particle Green function -- to Cooper pairs,
and then we get the result for Bose--condensate as in GL theory. 3)
Finally, the interaction (contribution of fourth order) appears in
GL theory due to the thing that Cooper pairs overlap, i.e. because
of Pauli principle. If it does not happen, i.e. they are separated,
then there is no interaction. It is not seen in the traditional
approach because self--action always remains.

Note that the size of a pair does not change with temperature on the
order of value, whereas their number decreases. Therefore, one can
consider the separated pairs near the transition point.

The presented arguments are a ground for the suggested
generalization. It is worth emphasizing that the description of
superconducting state as a set of particles (Cooper pairs) -- as it
is proposed -- appears to be natural, but it is not strictly
grounded and, thus, it is no more than a model.

\section{Discussion of the model}

Let a superconducting plate be in the transverse magnetic field H.
Abrikosov vortex lattice forms  near the field $H_{c2}$. The lattice
can melt analogously to any crystal melting. When discussing melting
(see e.g. Review [3]), vortices are thought to remain, but the long
range order of their arrangement disappears. It turns out that a
liquid state of other type is possible near $H_{c2}$ and it is
different from all considered earlier [3]. Namely, there are no
vortices but there is a homogeneous system -- liquid of Cooper pairs
(without Bose--condensate and vorticies). Such a picture naturally
appears in Cooper pairs description assisted with the Hamiltonian
(2).

First it is necessary to determine what to consider as particles
(Cooper pairs) concentration. It would be necessary to think
$|\Psi|^2$ as such one, i.e. without external fields
$|\alpha|/\beta$ (in the unit of volume of a 3D case). However, in
GL theory, this value is not determined because any value, different
from $\Psi$ in its constant multiplier, may be taken as an order
parameter, lest the physical values should not be unchanged. These
values are: 1) magnetic field penetration depth \begin{equation}
\label{3} \lambda(T) = \sqrt{\frac{Mc^2\beta}{4\pi(2e)^2|\alpha|}}\
,\end{equation}  2) length of coherence \begin{equation} \label{4}
\xi(T)=\frac{1}{\sqrt{2M|\alpha|}}\ ,\end{equation} 3) free energy
change under the transition into ordered state
\begin{equation} \label{5} \frac{\delta
F}{V}=-\frac{\alpha^2}{2\beta} \rightarrow -\frac{A^2\gamma
T_c^2}{2}(1-T/T_c)^2\end{equation} (the presented value is from the
microscopic theory, $\gamma$ is density of states on Fermi surface,
$A\approx 3.06$). It is seen from these expressions that only two
were fixed (all is clear with the charge), out of these three GL
theory parameters $\alpha,\beta,M$. As for the rest, there is
arbitrariness and ratio $|\alpha|/\beta$ remains indefinite, which
is irrelevant for the GL approach. It is common knowledge.

In the operator formulation (2), when we say about particles (Cooper
pairs) and their quantity, it is necessary to register the left
parameter $|\alpha|/\beta$ which just corresponds to particles
density. For this purpose it is necessary to estimate the number of
Cooper pairs $N_{CP}$. Let us show the way it is possible to make
it.

Consider a usual superconductor. Here it is clear how to estimate
it. First, let us present some known data from the theory of
superconductivity which are necessary for estimations.  In
Bardeen--Cooper--Schrieffer model (only electrons attraction with
opposite momenta and spins are considered in the Hamiltonian) , the
transition from electron operators $a_{{\bf p}\sigma}$ to
quasi--particles operators $\alpha_{{\bf p}\sigma}$ is realized by
Bogolyubov transformations:
\begin{eqnarray} \label{6} a_{{\bf p}\uparrow} =u_{\bf p}\alpha_{{\bf
p}\uparrow} +v_{\bf p} \alpha^+_{{-\bf p}\downarrow}\ ,\ \ a_{-{\bf
p}\downarrow} =u_{\bf p}\alpha_{-{\bf p}\downarrow} -v_{\bf p}
\alpha^+_{{\bf p}\uparrow}\ ,\end{eqnarray}
$$(u_{\bf p}^2,v_{\bf p}^2) = \frac{1}{2}\Biggl(1\pm\frac{\xi_{\bf p}}{\epsilon_{\bf p}}
\Biggr)\ ,\ \ \ u_{\bf p}v_{\bf p}=\frac{-\Delta}{2\epsilon_{\bf
p}}\ ,$$ where $ \epsilon_{\bf p} =\sqrt{\xi_{\bf p}^2 +\Delta^2}$
is quasi--particles spectrum, $\xi_{\bf p}$ is electron energy
calculated from Fermi energy.

Consider the mean value  $<\Psi_\uparrow({\bf
r})\Psi_\downarrow({\bf r}')>$. It is proportional to the electron
wave function in a pair. If we calculate the double--integral from
the square of this value's module, then there will be the number of
electrons in pairs (with the spin upwards and down), i.e. pairs
doubled number. Thus, \begin{eqnarray} \label{7}
N_{CP}(T)=\frac{1}{2}\int d{\bf r}d{\bf r}'\bigl|<\Psi_\uparrow({\bf
r})\Psi_\downarrow({\bf r}')>\bigr|^2\ .\end{eqnarray}Here,
operators $\Psi_{\uparrow,\downarrow}$ are usual field electron
operators, e.g.: $$\Psi_\uparrow({\bf r}) =
\frac{1}{\sqrt{V}}\sum_{\bf p}a_{\bf p\uparrow}\exp(i{\bf pr})\ .$$

Using the transition to quasi--particles operators and calculating
the mean value, we have:  $$<\Psi_\uparrow({\bf
r})\Psi_\downarrow({\bf r}')> = \frac{1}{V}\sum_{\bf
p}\frac{\Delta}{2\epsilon_{\bf p}} \tanh\biggl\{\frac{\epsilon_{\bf
p}}{2T}\biggr\}\times$$
$$\times\ \exp\Bigl[i{\bf p}({\bf r-r'})\Bigr] .$$ Integration on co--ordinates in value
$N_{CP}(T)$ gives multiplier $V^2 \delta_{\bf p,q}$ (${\bf p,\ q}$
are momenta in different sums); so, finally we will get:
\begin{eqnarray} \label{8} N_{CP}(T)=\frac{1}{2}\sum_{\bf
p}\frac{\Delta^2}{4\epsilon_{\bf
p}^2}\tanh^2\biggl\{\frac{\epsilon_{\bf p}}{2T}\biggr\}\
.\end{eqnarray} At zero temperature we have:
$$N_{CP}(0)/V=\frac{\gamma \Delta_0}{2}\ \frac{\pi}{4}\ .$$
Near the transition point there is: \begin{eqnarray} \label{9}
\frac{|\alpha|}{\beta}\
\equiv\  N_{CP}(T\approx T_c)/V = \frac{\gamma\Delta^2}{8T_c}\ D ;\\
\nonumber D\equiv\int_0^\infty \frac{dx}{x^2}\ \tanh^2(x)\approx
1.7\ .\end{eqnarray}

As is known the gap in the spectrum near transition point is
\begin{equation} \label{10} \Delta= A\ T_c \sqrt{1-T/T_c}\
.\end{equation} This value, just as ratio $\alpha^2/\beta$ (see
(5)), does not change in dirty superconductor (Anderson theorem).
Besides, the expression  for pairs number (9) remains the same. It
turns out that, with impurities, only mass $M$ changes (see below
(12)).

It is not difficult to get mass $M_0$ (for a pure case) assisted
with expressions for other constants. For instance, for $\lambda(T)$
(see (3)) we have: \begin{equation} \label{11} \lambda(T)
=\frac{\lambda(0)}{\sqrt{2(1-T/T_c)}}\ ,\ \ \ \ \
\lambda(0)=\sqrt{\frac{mc^2}{4\pi ne^2}}\ ,\end{equation} where the
known ratio of superfluid component near the transition point for
pure superconductor ($m$ -- electron mass, $n$ -- concentration) is
used. It is sufficient for our purposes. As  a result:
\begin{equation} \label{12} \frac{M_0}{m}=\frac{3A^2D}{16}\
\frac{T_c}{\epsilon_F}\ ,\ \ \ \ \ \
\frac{M}{M_0}=\frac{\kappa}{\kappa_0}\sim \frac{\xi_0}{l}\
(l<\xi_0)\ . \end{equation} Here $\kappa=\lambda(T)/\xi(T)$ is a GL
parameter ($k_0$ -- for pure sample), $l$ -- free path length. Let
us point out the thing that it is when Cooper pairs number is meant
by particles number.

Now one more expression for coherence length (see (\ref{3}),
(\ref{4}), (\ref{5}) и (\ref{11})): \begin{equation}
\label{13}\xi(T) =\frac{1}{A\sqrt{6}}\
\frac{v_F/T_c}{\sqrt{1-T/T_c}}\ .
\end{equation} The last three expressions are for pure superconductor (not
considering the expression for $M$).

Finally, let us put down the expression for coefficient $\beta$
which is derived from comparison (5) and (9) using (10):
\begin{equation} \beta =\biggl(\frac{8}{A} \biggr)^2\
\frac{1}{\gamma D^2}\ . \label{14}\end{equation}

Using expression (9), it is possible to get the evaluation of
temperature interval near the transition point $\delta T/T_c$ when
the mean distance between Cooper pairs becomes larger than the size
of Cooper pair $\xi_0$. For pure superconductor $\xi_0\sim v_F/T_c$,
so it makes: \begin{equation} \label{15} \frac{\delta T}{T_c}\sim
\biggl(\frac{T_c}{\epsilon_F}\biggr)^2\ .\end{equation} It is in
this region that one can consider Cooper pairs as particles; though
this interval is still beyond the limits of fluctuation region which
occurs near the transition point in the interval $\delta
T/T_c\sim(T_c/\epsilon_F)^4$ but it is hardly possible to see
anything new here.

\section{New state}

The superconducting plate in transverse magnetic field is analyzed.
Let the thickness of plate $d$ is small enough -- less than the pair
size but more than the distance among electrons:
$$1/p_F<<d<\xi_0,\sqrt{\xi_0l}\ .$$

The last value is the pair size in a dirty sample ($l$ - free path
length). In this case the above--mentioned estimates for a massive
sample are suitable but, instead of the Hamiltonian (2), it is
necessary to write down a two--dimensional equivalent. Namely:
\begin{eqnarray} \nonumber \textsf{H} \rightarrow  \int d{\bf
r}\biggl\{\frac{-1}{2M} \Psi^+({\bf r}) \Bigl(\nabla - i\frac{2e}{c}
{\bf A}({\bf r})\Bigr)^2\Psi({\bf r}) +\\ \alpha \Psi^+({\bf
r})\Psi({\bf r}) + \label{16} \frac{\beta_2}{2} \Psi^+({\bf r})
\Psi^+({\bf r}) \Psi({\bf r})\Psi({\bf r})\biggr\}\
.\end{eqnarray}Two--dimensional integration (in the plate's plane)
is implied here and $\beta_2 \equiv \beta/d$, and all the operators
are two--dimensional.

We are focused on the basic state of particle (Cooper pair) in the
magnetic field. As is known, in the cylindrical calibration of
vector potential ($A_\varphi = H\rho/2,\ A_\rho=A_z=0$) wave
functions of the basic Landau level are registered as: $$\phi_n \sim
z^n\exp\biggl(-\frac{|z|^2}{4a_H^2}\biggr),\ \ \ \ z=x+iy\ .$$ Here
$n$ has non-negative integral values (n = 0, 1, 2...),  $a_H$ --
magnetic length (for Cooper pair $a^2_H = c/(2|e|H)$). Energy
\begin{equation} \label{17} E_1=\frac{1}{2}\ \frac{2|e|H}{Mc}\ ,\ \
\ \ \ \frac{|e|H_{c2}}{Mc}=|\alpha|\end{equation} corresponds to
these states.

It is possible to employ a Laughlin--like function [5] (functions of
this type were used in the theory of fractional quantum Hall effect)
-- naturally, regarding Bose--particles. The advantage of such
functions  consists in the thing that short--range interaction is
completely excluded, just as in the case we are focused on.
Laughlin--like function $\Phi_L$, in our case, is the following:
\begin{equation} \label{18}\Phi_L\ \sim\ \prod_{i>j=1}^N (z_i-z_j)^2 \prod_{k=1}^N
\exp\biggl(-\frac{|z_k|^2}{4a_H^2}\biggr)\end{equation} ($N$ --
number of Cooper pairs). This function is symmetrical in its
particles exchange, just as it is required, and the interaction
turns to zero. Note that it is possible to analyze any even power
instead of square, particles concentration being lower. For function
(18) the concentration is maximally possible (one particle per two
magnetic flux quanta), i.e. the energy is minimally lower near the
transition point ($H<H_{c2}$). This is the state with unconsidered
fluctuations (without other Landau levels).
   As a result, the mean value of operator (16) on state (18) is equal to:
\begin{eqnarray} \label{19} \frac{<\textsf{H}>}{S}
= -|\alpha| \biggl( 1-\frac{H}{H_{c2}}\biggr)\frac{|e|H}{2\pi
c}\approx \\ \nonumber \approx\  -|\alpha| \biggl(
1-\frac{H}{H_{c2}}\biggr)\frac{|e|H_{c2}}{2\pi c}\ ,\end{eqnarray}
where $S$ is area. It is taken into account that the interaction in
this state is equal to zero.
   For Abrikosov vortex state we have:  \begin{equation}
\label{20} \frac{F_S- F_N}{S} = \frac{-\alpha^2}{2\beta_2 C} \biggl(
1-\frac{H}{H_{c2}}\biggr)^2 \ . \end{equation} Here constant $C$
depends on lattice type (for triangular lattice $C\approx 1.16$).

Thus, one should compare two energies: $$\frac{-\alpha^2}{2\beta_2
C} \biggl( 1-\frac{H}{H_{c2}}\biggr)^2\ \div\ -|\alpha| \biggl(
1-\frac{H}{H_{c2}}\biggr)\frac{|e|H_{c2}}{2\pi c}\ .$$ This or that
state is realized depending on what energy is lower. Whence forth,
there is the magnetic field interval when the state occurs according
to Laughlin: \begin{equation}\label{21}\frac{\delta H}{H_{c2}}=
\frac{M\beta}{d}\ \frac{C}{\pi}\ .\end{equation}  Using the previous
result we have: \begin{equation}\label{22} \biggl(\frac{\delta
H}{H_{c2}}\biggr)_0=\frac{T_c}{\epsilon_F}\ \frac{1}{p_Fd}\
\frac{24\pi C}{D}\approx\frac{T_c}{\epsilon_F}\ \frac{51.448}{p_Fd}
\ .\end{equation} Index $0$ means that this expression is for a pure
sample (mirror walls of a plate and no impurities). It is a small
value for a plate from usual superconductors even with a large
numerical coefficient.

In case of a dirty sample (free path length is small $l<<\xi_0\
,d$), there occurs an additional multiplier:
\begin{equation}\label{23}\frac{\delta
H}{H_{c2}}=\biggl(\frac{\delta H}{H_{c2}}\biggr)_0\
\frac{\kappa}{\kappa_0}\sim\biggl(\frac{\delta H}{H_{c2}}\biggr)_0\
\frac{\xi_0}{l}\end{equation} (see expression (12)). By the by, the
upper critical field $H_{c2}$ increases also:
$$\frac{H_{c2}}{(H_{c2})_0}=\frac{\kappa}{\kappa_0}\ .$$ Conclusion:
the thinner and dirtier the plate is, the larger the interval of a
new state's existence is.

Thus, in the interval of fields (21) (see also (22), (23)) there is
a liquid state (incompressible liquid, as such a state is called in
the theory of fractional quantum Hall effect). This state is
explicitly different from Abrikosov vortex lattice. Even if the
lattice melted, there still is a transition into a new state because
the dependence of energy from the value $(1 - H/H_{c2})$ is linear,
whereas this dependence remains quadratic, as in (20), for the
melted state of Abrikosov lattice, although with other coefficients
(the same on value order).

Both transitions on the edges of the interval are transitions of the
fist kind with magnetization hops and magnetization is constant
inside the interval.

Some things about high-temperature superconductors. As they are a
set of weakly connected 2D layers, one can expect that the state
discussed here is also possible for them. If it is so, then the
corresponding magnetic field interval is to be significantly larger
(temperature of transition $T_c$ is higher, thickness of layer $d$
is smaller).

My thanks to A.V. Chaplik and M.V. Entin for discussion.

\end{document}